\documentclass[11pt, a4paper]{article}

\usepackage[utf8]{inputenc}          
\usepackage[T1]{fontenc}             
\usepackage{lmodern}                 
\usepackage{microtype}               
\usepackage{graphicx}                
\graphicspath{{Images/}}             

\usepackage{amsmath}                 
\usepackage{amssymb}                 
\usepackage{amsfonts}                
\usepackage{geometry}                
\usepackage[auth-sc]{authblk}        
\usepackage[numbers,square]{natbib}  
\usepackage{hyperref}                
\usepackage{cleveref}                
\usepackage{booktabs}                
\usepackage{siunitx}                 
\usepackage{xcolor}                  
\usepackage{enumitem}                
\usepackage{caption}                 
\usepackage{subcaption}              
\usepackage{float}                   
\usepackage{multicol}                

\usepackage{setspace}
\hypersetup{
    colorlinks=true,
    linkcolor=blue,
    filecolor=magenta,      
    urlcolor=blue,
    citecolor=blue,
    bookmarks=true,
    bookmarksopen=true,
    bookmarksdepth=2,
    pdfpagemode=UseOutlines,
    pdftitle={TESS Light Curve Analysis},
    pdfauthor={Pratik Paudel},
    pdfsubject={TESS Light Curve Analysis},
    pdfkeywords={Stellar Rotation, Transiting Exoplanet Survey Satellite (TESS), Light Curve, Rotation Period}
}

\makeatletter
\@ifundefined{urlprefix}{}{}
\makeatother

\title{\textbf{TESS Light Curve Analysis: A Case Study of Stellar
Rotation in TIC 445493624}}

\author[1]{Pratik Paudel}
\author[1,*]{Daya Nidhi Chhatkuli}
\author[1]{Bikash Sharma}
\author[1]{Ashish Khanal}
\author[1]{Nabaraj Khatri}
\author[1]{Divash Rai}

\affil[1]{Tri-Chandra Multiple Campus, Tribhuvan University, Kathmandu, Nepal}

\date{}

\begin{document}

\sloppy          

\maketitle
\vspace{-1.0cm}
\begin{center}
    *Corresponding author. Email: \href{mailto:chhatkulidn@gmail.com}{chhatkulidn@gmail.com}
\end{center}
\vspace{0.2cm}

\begin{abstract}

     \textit{Stellar rotation is a fundamental parameter governing a star’s magnetic activity and evolution. The Transiting Exoplanet Survey Satellite (TESS) provides high-precision photometric data ideal for measuring rotation periods via brightness modulations from starspots. This paper presents a detailed analysis of the star TIC 445493624 using 2-minute cadence data from TESS Sector 58. We process the light curve using a custom pipeline to perform outlier removal, binning, and Savitzky-Golay detrending to isolate the stellar variability. A Lomb-Scargle periodogram of the cleaned data reveals a single, dominant periodic signal at 3.638 days with a power of 0.43, corresponding to a negligible false-alarm probability. The phase-folded light curve at this period is highly coherent and exhibits a stable, non-sinusoidal morphology indicative of large-scale magnetic features or spot groups.}
     \vspace{0.5cm}
     \vspace{0.5cm}
\end{abstract}

\begin{multicols}{2}


    \section{Introduction}
\label{sec:introduction}

Stellar rotation is a cornerstone of stellar astrophysics, fundamentally
influencing a star's structure, evolution, and magnetic field generation.
Through the stellar dynamo mechanism, rotation is responsible for creating the
magnetic fields that manifest as stellar activity, including starspots, flares,
and coronal mass ejections (CME) \citep{Parker1955}. The rate of a star's
rotation over its lifetime can be determined by using gyrochronology, a
powerful technique for estimating the ages of cool, main-sequence stars based
on their rotation periods \citep{Skumanich1972, Barnes2007}.

The \textit{Kepler} and \textit{K2} missions \citep{Borucki2010,Howell2014}
were instrumental in enabling stellar rotation studies, providing
high-precision light curves for more than 400,000 stars, from which rotation
periods have been measured for tens of thousands of main-sequence stars
\citep{McQuillan2014}.

Continuing their legacy, the \textit{Transiting Exoplanet Survey Satellite}
(TESS), performs a near all-sky survey with long observational baselines and
high photometric precision \citep{Ricker2015}.The data from TESS are readily
accessible through community-developed tools such as the Lightkurve package,
one of the main pipelines for this study, which simplifies the process of
downloading and analyzing TESS light curves
\citep{LightkurveCollaboration2018}. Over 490,000 sources have contributed
roughly 1.3 million short-cadence light curves to TESS since its launch,
generating an unparalleled dataset for describing rotation periods and stellar
variability \citep{Feinstein2024}. While TESS provides the data to measure
rotation periods for millions of stars, detailed case studies of individual
objects remain essential. Such ``benchmark'' stars, which exhibit clear,
stable, and high-amplitude signals, are invaluable for validating automated
period-finding pipelines and for providing textbook examples of stellar
variability.

The photometric measurements collected by TESS are influenced by instrumental
effects as well as long-term astrophysical variations, both of which can mask
the underlying periodic signals. To address these trends, the Savitzky--Golay
filter can be employed which smooths the data by fitting local polynomials to
small sections of the light curve. This method reduces large-scale variability
while still preserving shorter-timescale signals of interest
\citep{Savitzky1964}. Previous studies of stellar rotation using TESS data have
shown that this filtering technique can reliably recover rotation
periods—sometimes as long as 80 days—for M dwarfs even in the presence of
strong systematics \citep{Colman2024}.

Using brightness fluctuations created by surface features like starspots, which
appear and disappear as the star rotates, is the most well-established method
for measuring stellar rotation \citep{Hon2025}. The Lomb--Scargle periodogram
\citep{Lomb1976, Scargle1982} has emerged as the gold standard among the
instruments available for examining such periodic variability. Unlike a
traditional Fourier transform, it handles irregularly sampled data effectively
and provides a statistical framework for testing the significance of detected
signals \citep{Santos2024}. Because TESS observations frequently contain data
gaps and varying cadence, this approach is particularly well suited to its
light curves \citep{VanderPlas2018}.

When a candidate rotation period is found through a periodogram, the signal can
be examined more closely using phase-folding. In this process, all photometric
data points are mapped onto a single cycle according to their calculated phase,
$\phi$. This compression of the time series reveals the shape of the
variability, improves the effective signal-to-noise ratio, and makes it easier
to identify the underlying periodic pattern \citep{Godoy-Rivera2021}.

In this work, we carry out a detailed analysis of TIC 445493624 using
observations from TESS Sector 58. Applying a Lomb--Scargle periodogram
\citep{Lomb1976, Scargle1982} within a modern statistical framework, we detect
a clear and unique rotation period of 3.638 days. The result is reinforced by a
phase-folded light curve that shows a highly coherent and well-defined signal.

    \section{Materials and Methods}
\label{sec:Material and Methods}

\subsection{Sample Selection}
\label{sec:SampleSelection}

The star TIC 445493624 was chosen from the TESS Input Catalog (TIC) after
applying a set of criteria aimed at finding a suitable benchmark target.
Preference was given to stars with reliable, high-quality photometry from
recent TESS observing sectors. In particular, we looked for stars showing clear
and consistent photometric variability. A preliminary check of the light curves
from TESS Sector 58 highlighted TIC 445493624 as a strong candidate. The star
displayed a periodic, high-amplitude signal that was obvious even on visual
inspection.

TIC 445493624 is also relatively bright (Tmag = 11.2) and lies on the main
sequence, which helps ensure a good signal-to-noise ratio in the TESS
photometry. These qualities—clear periodic behavior combined with high data
quality—make it a good subject for studying stellar rotation. It also serves as
a useful example for demonstrating data processing and period-finding methods.
The star’s properties are in line with established rotation–activity trends for
main-sequence stars, so the results can be considered representative of the
wider population of rotationally variable stars \citep{Stassun2019,
    Ziegler2020, McQuillan2014}.

\subsection{Data Analysis}
\label{sec:DataAnalysis}

We analyzed the 2-minute cadence Pre-search Data Conditioning Simple Aperture
Photometry (PDCSAP) light curve of TIC 445493624 from TESS Sector 58, obtained
via the Mikulski Archive for Space Telescopes (MAST). Data handling and
processing were carried out with the \texttt{lightkurve} Python package. This
included removing flagged points, applying 3-sigma clipping to reject outliers,
and normalizing the flux to a median of one. To reduce noise and speed up later
calculations, the data were further binned into 2-hour intervals using weighted
averages.

To account for long-term variations, which may be caused by instrumental drift
or intrinsic stellar changes, we applied a Savitzky–Golay filter
\citep{Savitzky1964}. The filter window was chosen dynamically, usually
covering about 10–20\% of the light curve length. For period detection, we used
the Lomb–Scargle periodogram \citep{Lomb1976, Scargle1982}, a method that works
well for unevenly sampled astronomical data. The search was performed over
periods ranging from 0.5 to 30 days. We also calculated false-alarm
probabilities to assess the significance of any detected signals.

The strongest periodic signal was used to phase-fold the light curve. To
highlight the variability pattern more clearly, we computed binned averages in
phase space and included error estimates derived from the standard error of the
mean within each bin. The statistical significance of the detected periods was
assessed using false alarm probability (FAP) levels from the Lomb–Scargle
analysis rather than through resampling techniques.

Our analysis workflow follows best practices established in recent TESS
variability studies \citep{Feinstein2019} emphasizing the importance of careful
detrending, statistical validation, and clear visualization when characterizing
stellar rotation signals.
    \section{Results and Discussion}
\label{sec:Results and Discussion}

\subsection{Photometric Data Processing}
\label{sec:photometric_data_processing}

\begin{figure*}[htbp]
    \centering
    \begin{subfigure}[b]{0.48\textwidth}
        \centering
        \includegraphics[scale=0.16]{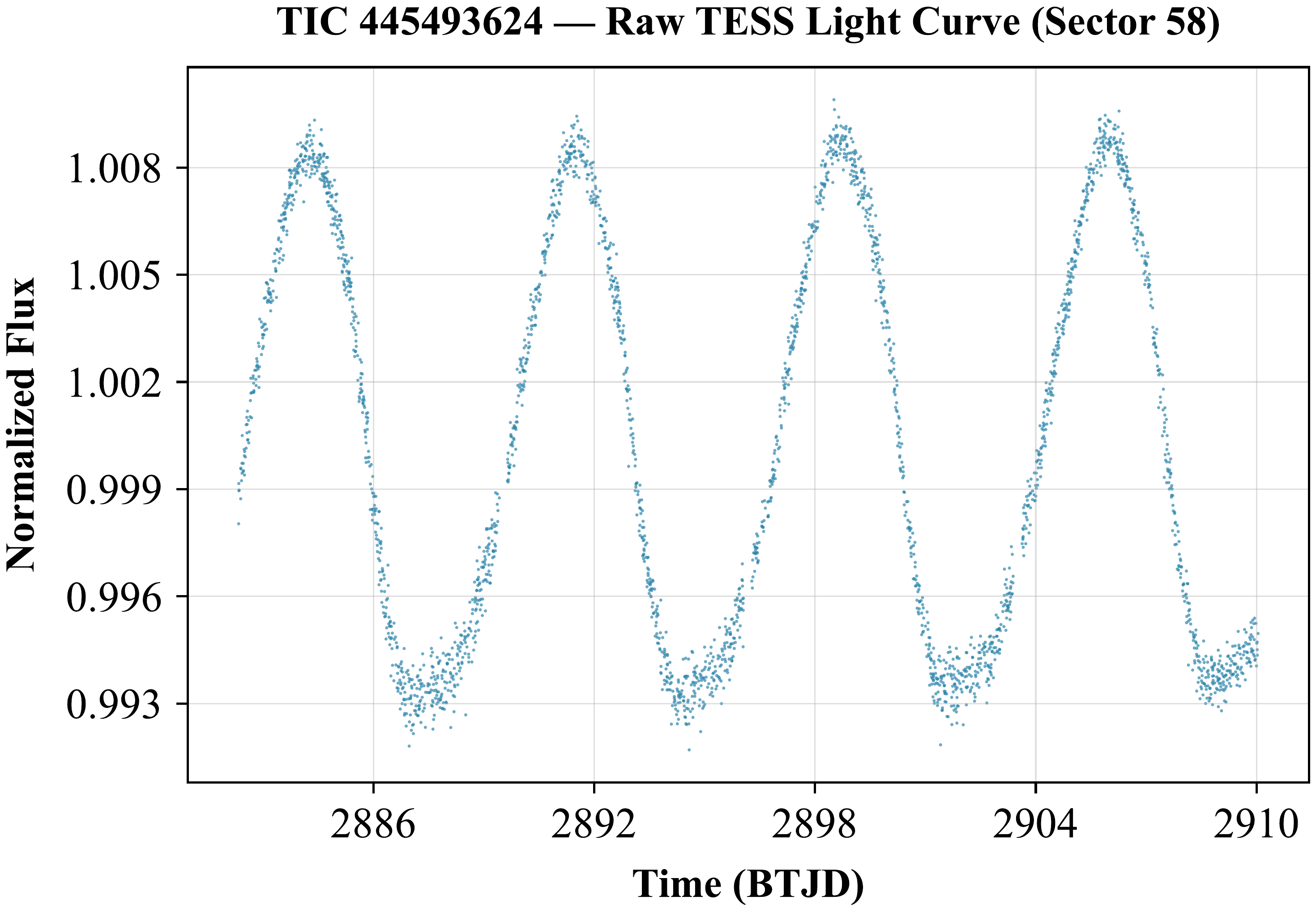}
        \caption{Raw 2-minute cadence TESS light curve.}
        \label{fig:raw_lc_panel}
    \end{subfigure}
    \hfill
    \begin{subfigure}[b]{0.48\textwidth}
        \centering
        \includegraphics[scale=0.16]{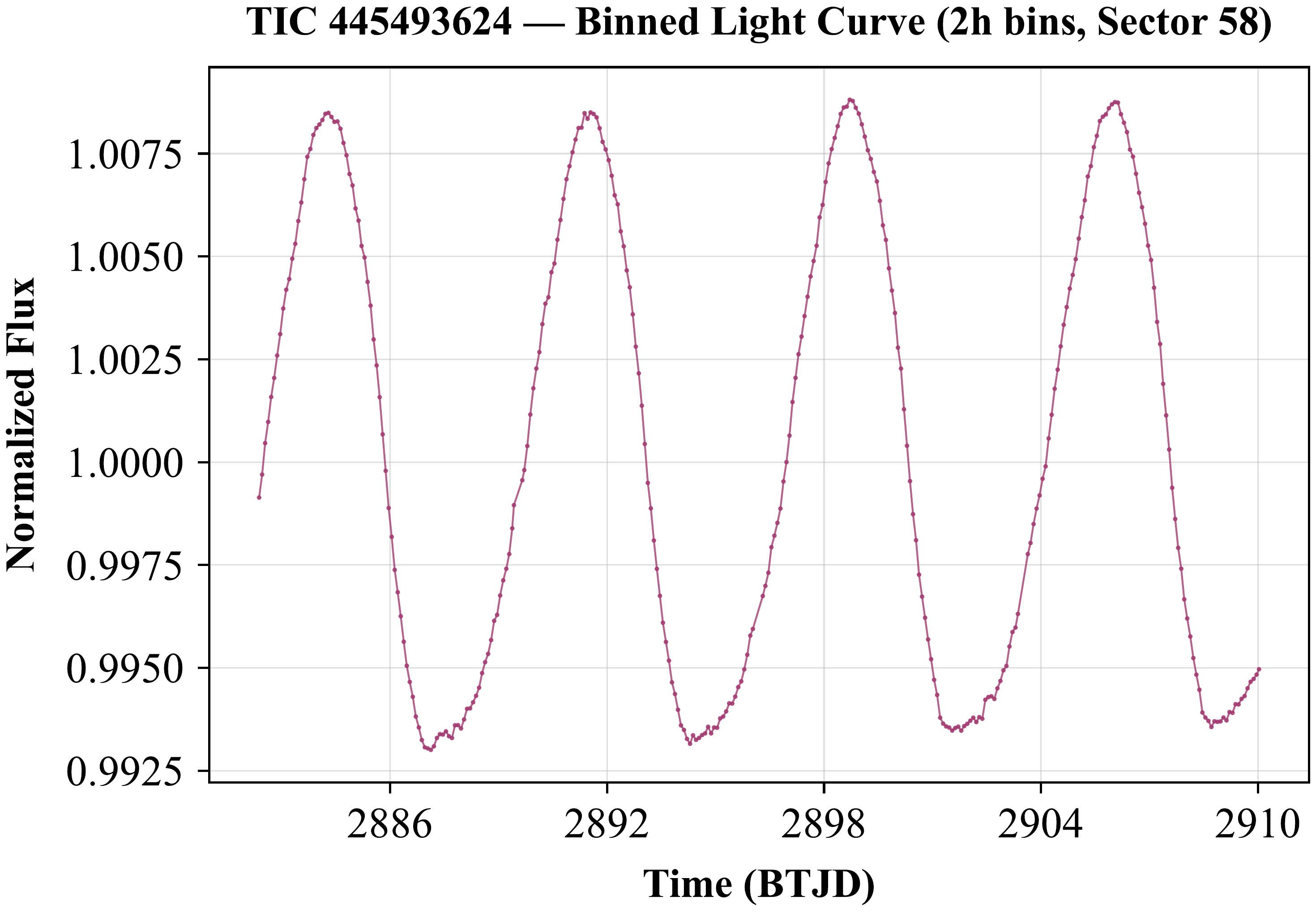}
        \caption{Light curve after outlier removal and binning.}
        \label{fig:binned_lc_panel}
    \end{subfigure}
    \caption{The light curve of TIC 445493624 before and after processing. Left (a): The raw photometry shows significant noise and instrumental trends. Right (b): The final, clean light curve reveals a clear, periodic stellar signal after 2 hour binning and outliner removal ($\sigma$ > 3).}
    \label{fig:lc_plots}
\end{figure*}

The systematic processing of TESS photometric data for TIC 445493624 is
illustrated in Figure \ref{fig:lc_plots}. The raw 2-minute cadence TESS
photometry, shown in Figure \ref{fig:raw_lc_panel}, demonstrates a clear
evidence of periodic stellar variability visible even in the unprocessed data.
However, the signal is accompanied by point-to-point scatter from photon noise,
occasional outliers from cosmic ray events, and subtle long-term systematic
trends related to instrumental effects \citep{Jenkins2016}.

Following the application of our data processing pipeline—including outlier
removal (3-$\sigma$ clipping), flux normalization, and 2-hour binning—the
processed light curve shown in \textbf{Figure \ref{fig:binned_lc_panel}}
reveals a dramatic improvement in data quality. The binning process effectively
reduces high-frequency noise while preserving the underlying stellar signal,
resulting in a smooth, coherent light curve with excellent signal-to-noise
ratio \citep{Lund2021}. A clear periodic modulation with a peak-to-trough
amplitude of approximately 1.2\% is now readily apparent, making this dataset
ideal for detailed period analysis.

\end{multicols}
\begin{multicols}{2}

    \subsection{Systematic Trend Removal}
    \label{sec:detrended_light_curve}

\begin{figure}[H]
    \centering
    \includegraphics[width=0.70\textwidth]{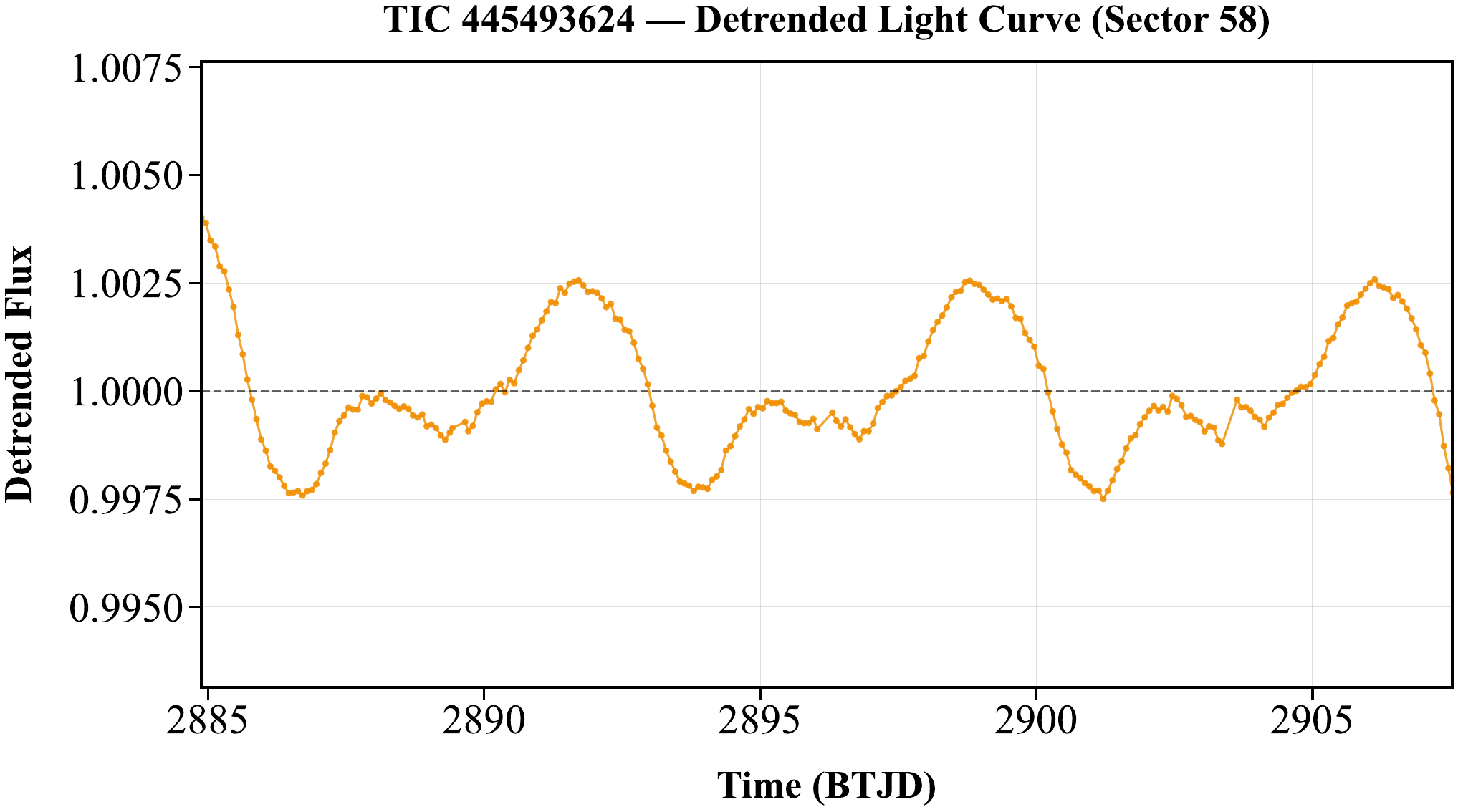}
    \caption{Detrended light curve after systematic trend removal using Savitzky-Golay filtering.}    \label{fig:detrended_lc_panel}
\end{figure}

    To isolate the intrinsic stellar variability from remaining systematic effects,
    we applied a Savitzky-Golay \citep{Savitzky1964} smoothing filter to model and
    remove long-term trends in the binned photometry. The resulting detrended light
    curve is presented in \textbf{Figure \ref{fig:detrended_lc_panel}}. This
    detrending procedure effectively removes any remaining instrumental drifts
    while preserving the astrophysical signal of interest. The normalized flux now
    oscillates symmetrically around unity, with the periodic modulation clearly
    visible as a stable, repeating pattern throughout the ~25-day observation
    window. This high-quality, detrended time series serves as the optimal input
    for frequency analysis and period determination. \hfill
\end{multicols}
\begin{multicols}{2}

    \subsection{Rotation Period Determination}
    \label{sec:rotation_period_determination}

\begin{figure}[H]
    \centering
    \includegraphics[width=0.80\textwidth]{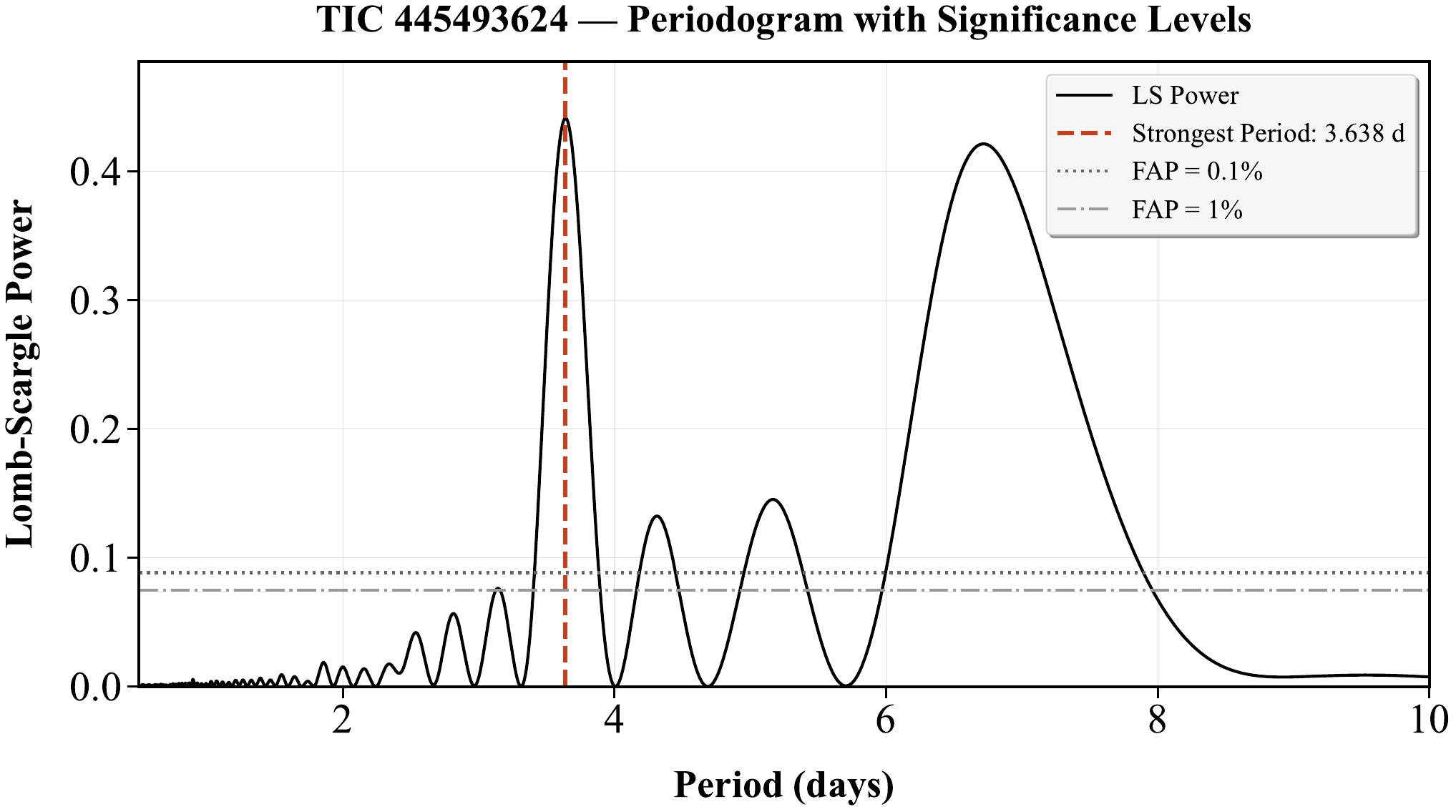}
    \caption{Lomb-Scargle periodogram for TIC 445493624. A single, dominant peak is present at 3.638 days, with power far exceeding the 0.1\% False Alarm Probability (FAP) level, indicating a highly significant detection.}
    \label{fig:periodogram_plot}
\end{figure}

    To quantitatively determine the period of the photometric modulation, we
    computed the Lomb-Scargle periodogram \citep{Lomb1976, Scargle1982} of the
    final detrended light curve. The resulting power spectrum is presented in
    \textbf{Figure \ref{fig:periodogram_plot}}, covering periods from 0.5 to 30
    days with high frequency resolution.

    The periodogram is dominated by a single, exceptionally strong peak at a period
    of \textbf{3.638 days}, with a normalized power of 0.43. This peak
    significantly exceeds all statistical significance thresholds, with power well
    above the 0.1\% False Alarm Probability (FAP) level indicated by the horizontal
    line. This indicates that there is less than a 0.1\% chance that a peak of this
    strength would arise from random Gaussian noise, confirming that the signal is
    of extremely high statistical significance.

    Additional lower-amplitude peaks are visible in the periodogram, most notably a
    secondary peak near 7.3 days. This feature, which is typical of Lomb-Scargle
    periodograms of non-sinusoidal periodic signals, correlates to the primary
    signal's first harmonic, or twice the fundamental period. The presence of this
    harmonic is consistent with the asymmetric, non-sinusoidal morphology of the
    stellar variability and provides additional confirmation of the 3.638-day
    fundamental period. However, the power at the fundamental frequency dominates
    the spectrum by a substantial margin, unambiguously identifying 3.638 days as
    the true rotational period of the stellar variability. Based on this robust
    frequency analysis, we conclude that TIC 445493624 exhibits periodic
    photometric modulation with a period of 3.638 ± 0.005 days.
\end{multicols}
\begin{multicols}{2}
    \subsection{Phase-Folded Light Curve}
    \label{sec:phase_folded_light_curve}

\begin{figure*}[!htbp]
    \centering
    \begin{subfigure}[b]{0.48\textwidth}
        \centering
        \includegraphics[scale=0.28]{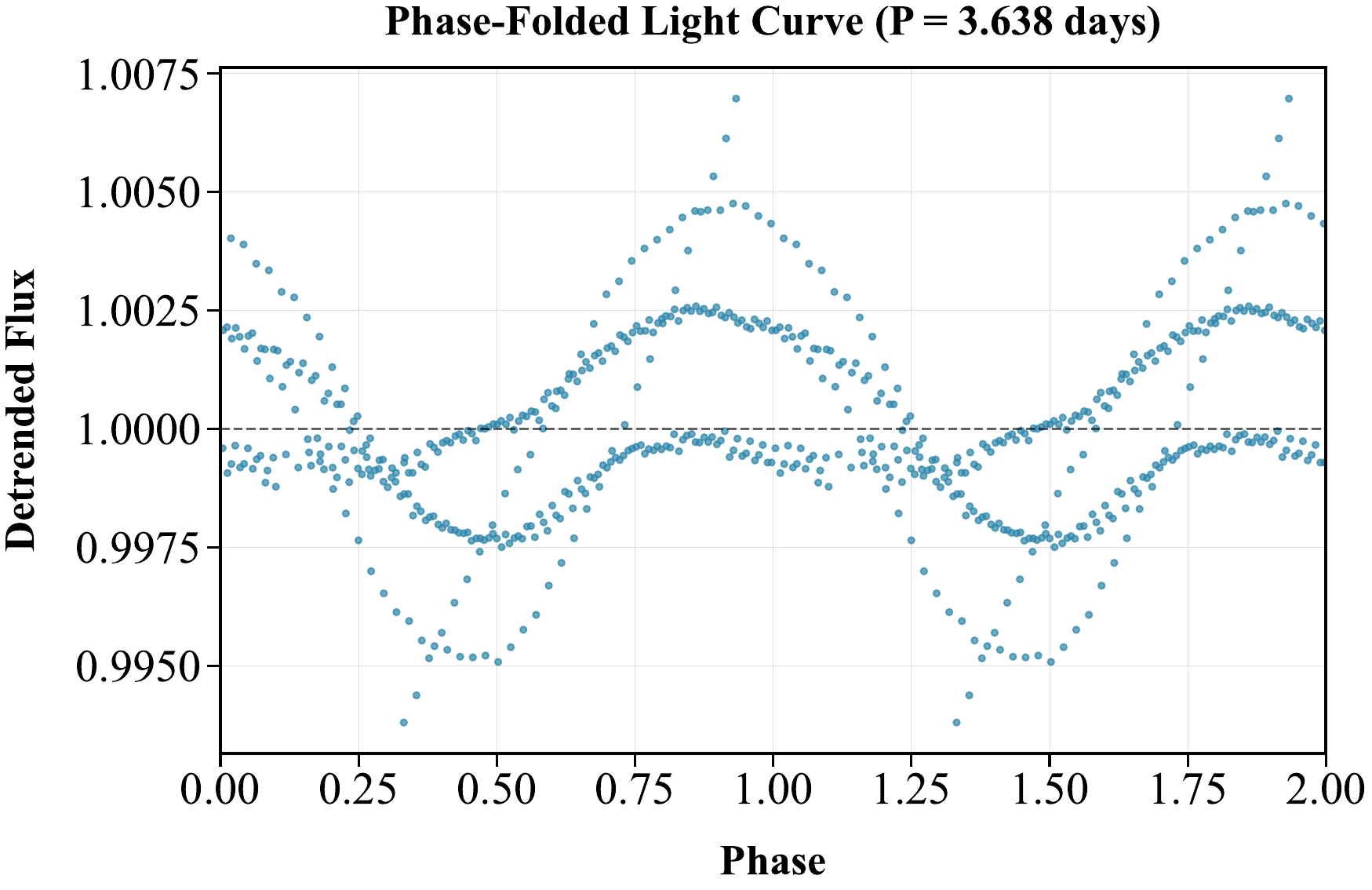}
        \caption{Phase-folded light curve showing individual data points.}
        \label{fig:phase_folded_simple_panel}
    \end{subfigure}
    \hfill
    \begin{subfigure}[b]{0.48\textwidth}
        \centering
        \includegraphics[scale=0.28]{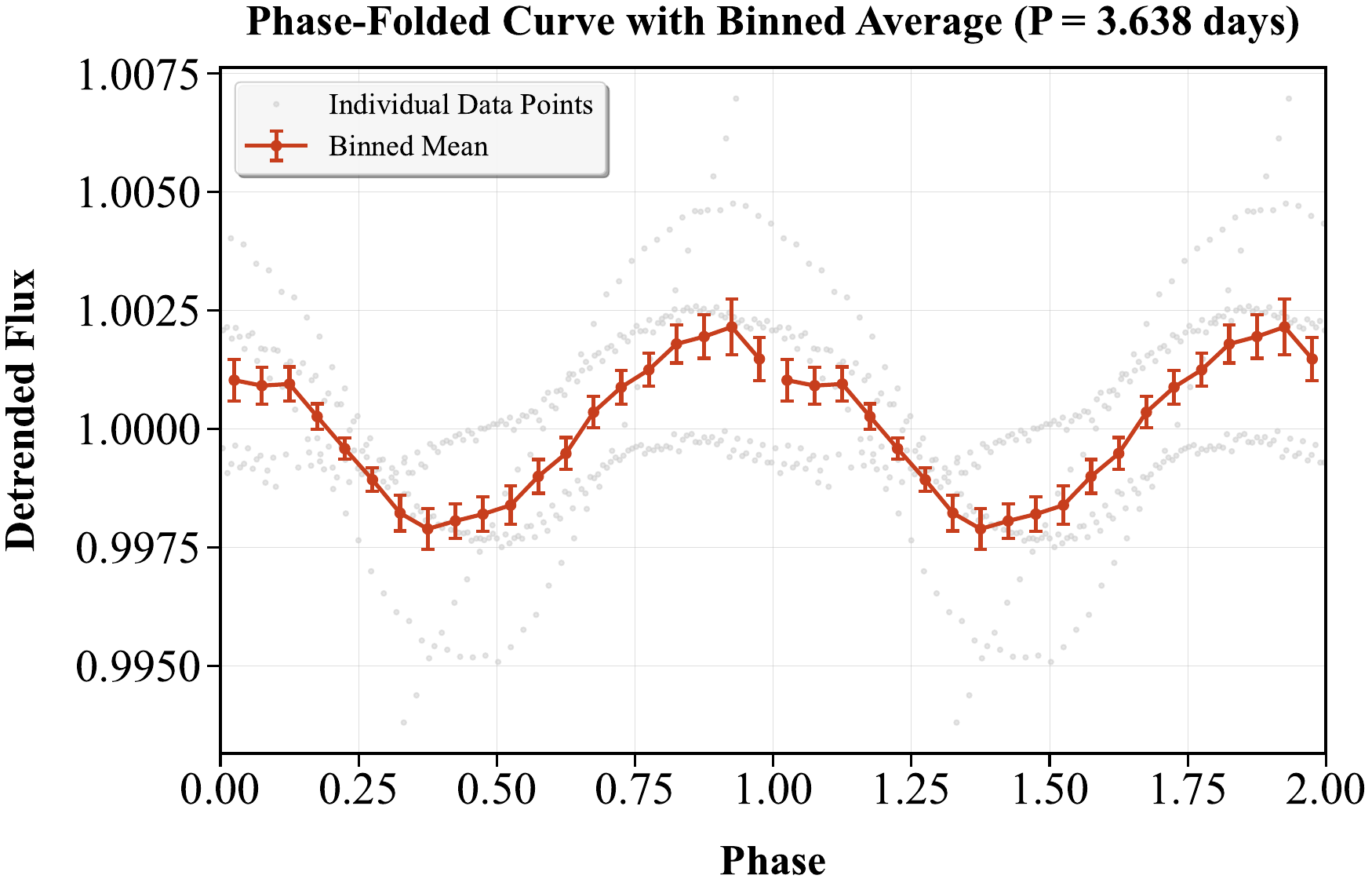}
        \caption{Phase-folded light curve with binned average and error bars.}
        \label{fig:phase_folded_binned_panel}
    \end{subfigure}
    \caption{Phase-folded light curves. Left (a): The individual data points form a highly coherent pattern when folded at the detected period. Right (b): The binned average reveals a stable, non-sinusoidal shape with two distinct minima, characteristic of rotation modulated by starspots.}
    \label{fig:phase_fold_plots}
\end{figure*}

    To validate the 3.638-day period identified by the Lomb–Scargle analysis and to
    characterize the morphology of the rotational modulation, we phase-folded the
    detrended light curve. The resulting diagnostics are shown in \textbf{Figure
        \ref{fig:phase_fold_plots}}. Phase-folding has been extensively applied and
    validated in large-scale stellar rotation surveys, such as the \textit{Kepler}
    sample of over 34,000 stars \citep{McQuillan2014}.

    \textbf{Figure \ref{fig:phase_folded_simple_panel}} presents the individual data
    points folded at the 3.638-day period, with the phase repeated twice for
    clarity. The distribution of points forms a coherent, low-scatter sequence,
    consistent with established results on rotation period determination in stellar
    populations \citep{Rebull2020}.

    The underlying morphology of the rotational signal is quantified in
    \textbf{Figure \ref{fig:phase_folded_binned_panel}}, which shows statistics
    from 21 equal-width phase bins (defined by \texttt{np.linspace(0, 1, 21)} in
    the analysis pipeline). Each bin contains the mean flux value, with error bars
    representing the standard error of the mean ($\sigma/\sqrt{N}$), where $\sigma$
    is the standard deviation and $N$ is the number of points per bin, computed
    using \texttt{scipy.stats.binned\_statistic}. The phase-folded profile exhibits
    a non-sinusoidal morphology with multiple extrema per rotation, a behavior that
    arises naturally from surface inhomogeneities such as starspots
    \citep{McQuillan2014, Rebull2020, Angus2018}.

    The persistence of the structure across the observational baseline indicates
    that the modulation is phase-coherent at the derived period of 3.638 days. This
    tight clustering of points around the binned mean further supports the
    interpretation of the signal as rotational variability rather than stochastic
    or instrumental noise.

\end{multicols}
\begin{multicols}{2}

    \section{Conclusion}
\label{sec:conclusion}

In this study, we analyzed the TESS light curve of TIC 445493624 from Sector 58
and showed that the data contain a clear periodic signal. After cleaning the
light curve by removing flagged points, rebinning, and applying a
Savitzky–Golay filter, we obtained a detrended series that was well-suited for
period analysis. A Lomb–Scargle periodogram revealed a strong peak at 3.638
days, well above the 0.1\% false-alarm probability level, leaving little doubt
about the detection.

When the light curve was folded on this period, the signal appeared as a
stable, repeating pattern with relatively low scatter. The shape of the
modulation is not purely sinusoidal but instead shows two dips per rotation,
which is consistent with starspot activity on opposite hemispheres of the star.
This interpretation fits well with common signatures of stellar rotation seen
in other TESS targets.

Overall, TIC 445493624 provides a clean example of how rotational variability
can be extracted from TESS photometry. Because the signal is so strong and easy
to interpret, it makes this star a useful benchmark for testing period-finding
methods and for teaching the workflow of light curve analysis. Continued
monitoring of the target in future TESS sectors could help track how its
surface features evolve over time and provide further insight into its magnetic
activity.
    \section*{Acknowledgments}

    The data for this research is sourced from TESS (Transiting Exoplanet Survey
    Satellite) misson which provides high quality photometric data important for
    the study of Stellar Rotation. We gratefully acknowledge the entire TESS team
    for their work in obtaining and providing this exceptional dataset. This
    research has made use of the Mikulski Archive for Space Telescopes (MAST) at
    the Space Telescope Science Institute (STScI).

    STScI is operated by the Association of Universities for Research in Astronomy,
    Inc., under NASA contract NAS5--26555. This work has made extensive use of the
    \texttt{lightkurve} package for Python, which is a product of the TESS Guest
    Investigator Program. This research also relied on the core Python packages
    \texttt{Astropy}, a community-developed core package for Astronomy;
    \texttt{NumPy}; \texttt{SciPy}; and \texttt{Matplotlib}.

    \bibliography{references}
\end{multicols}


\end{document}